# Extend the waveband of high isolation ratio and narrow band asymmetric transmission based on grating-film-grating nanostructure

Ruihan Ma[1], Yuqing Cheng[1,*] and Mengtao Sun[1,†]

[1] School of Mathematics and Physics, University of Science and Technology Beijing, Beijing 100083, People's Republic of China

**Abstract:** A serial asymmetric transmission (AT) nanodevices based on the grating-film-grating (G-F-G) structure are proposed and studied. By showing the results of three different nanodevices as examples, it is proved that this kind of G-F-G nanostructure can achieve high isolation ratio AT at arbitrary wavelength within a certain range by designing the parameters of the structure. These three nanodevices can achieve high forward transmittivities of 0.69, 0.71, and 0.87 at the wavelengths of 714, 810, and 905 nm, respectively, and the isolation ratio between forward and backward transmittivities are all more than 10 dB. In addition, the effects of the incidence angle and lateral displacement of upper gratings on device performance were also investigated, revealing that optimal AT is achieved under normal incidence and zero displacement. The proposed G-F-G structure provides a solution for passive and easy to manufacture optical isolators with potential applications in optical communication and sensing systems.

## 1. Introduction

Nanodevices with asymmetric transmission (AT) properties can provide difference in transmissivities of light between forward and backward transmission. These devices have the potential for application in integrated photonic chips, photonic crystals, nonlinear optical devices, and metasurface devices [1-6]. In recent years, artificial intelligence and edge computing have increased the demand for high-speed and low-power optical interconnects [7, 8]. AT devices help build direction-sensitive neuromorphic optical computing architectures, enabling the orderly distribution of optical signals within chips. They also help control the flow of information [9]. Furthermore, AT devices are important in optical security and information encryption. They improve system confidentiality and resistance to interference [10]. By using direction-dependent transmission channels, these devices allow information to be transmitted or read only in specific directions. AT devices have been achieved through several methods [11-15]. Feng et al. developed a metal-silicon waveguide system that demonstrated transmissive light propagation at the wavelength of 1550 nm. Near-field scanning optical microscopy characterization revealed unidirectional mode conversion from symmetric to antisymmetric modes during backward propagation [16]. Ba et al. designed a narrowband AT device based on a metal-metal-metal (M-M-M) structure. This device had a forward transmissivity of 0.72 at 610 nm and a bandwidth of 6.7 nm [17]. Liu et al. proposed a high-contrast dual-frequency AT device. This device used bilateral composite metal gratings, and it reached the transmittivities of 0.67 and 0.82 with corresponding contrast ratios 0.997 and 0.99 at 627 nm and 1238 nm, respectively [18]. In a previous work, Cheng et al. proposed a hybrid metallic nanowaveguide which has asymmetric gratings on both input and output ports to achieve AT with multiple modes in the visible range [19].

---

[*] Email: yuqingcheng@ustb.edu.cn

[†] Email: mengtaosun@ustb.edu.cn

In this study, we employ a grating-film-grating (G-F-G) asymmetric grating structure (the same as M-M-M structure) to achieve high isolation ratio with narrowband. Three structures with different parameters are shown, i.e., with upper grating periods of 700 nm, 800 nm, and 900 nm, respectively. These three devices work at three different wavelengths, i.e., 714, 810, and 905 nm, respectively. This work proves that the G-F-G structure can be employed to design AT nanodevices at arbitrate wavelength within a certain range.

## 2. Method and structure

Finite difference time domain (FDTD) method is employed to simulate the optical properties of the structure, and to design and optimize the parameters. The structure of the AT nanodevice is shown in Fig. 1. It consists of $SiO_2$ substrate, lower Ag gratings, middle Ag films, and upper Ag gratings. Particularly, the parameters of the upper and lower gratings are different, leading to AT in transmissivity [20, 21]. In the simulation, the refractive index of $SiO_2$ is 1.45, and the one of Ag is from the experimental data of Ref. [22]. The incident light is x-polarized plane wave along z (backward) or -z (forward) direction, with incident angle of 0 unless otherwise specified. Period boundary conditions in x and y directions are employed. Fig. 1 shows the unit cells of the nanostructure for three difference parameter sets, which have been optimized based on the forward transmittivities and the isolation ratio of transmittivities ($IR_T$) between forward and backward transmission. Here, $IR_T$ in unit dB is defined as:

$$IR_T = 10 \times \log_{10}\left(\frac{T_F}{T_B}\right) \tag{1}$$

where $T_F$ and $T_B$ stand for the transmittivities of forward and backward transmission, respectively. These three structures work at three different wavelengths, the parameters of which are shown in Table 1. The sickness of the middle Ag film is $d = 20$ nm for all the three devices. The lower surface of the middle film is defined as z=0.

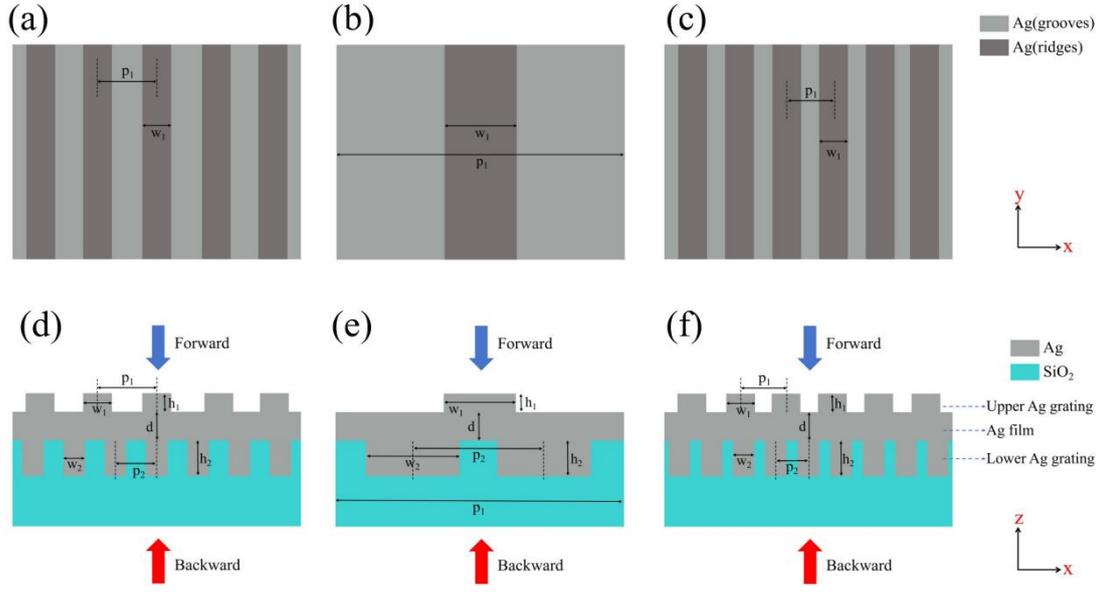

**Fig. 1.** (a)-(c) Top view of the $p_1 =$ 700, 800, and 900 nm unit cell. (d)-(f) Front view of the $p_1 =$ 700, 800, and 900 nm unit cell. The light gray areas represent the grooves, and the dark gray areas represent the ridges.

Table 1. Parameters and modes of the AT Devices

| Parameters | | unit: nm | |
| --- | --- | --- | --- |
| Upper grating period ($p_1$) | 700 | 800 | 900 |
| Lower grating period ($p_2$) | 500 | 400 | 600 |
| Upper groove depth ($h_1$) | 40 | 25 | 40 |
| Lower groove depth ($h_2$) | 150 | 180 | 150 |
| Upper ridge width ($w_1$) | 240 | 250 | 220 |
| Lower ridge width ($w_2$) | 220 | 290 | 200 |
| Resonance wavelength | 714 | 810 | 905 |
| FWHM | 12.3 | 4.2 | 3.5 |

## 3. Results and discussion

The working principle of AT in the proposed device relies on the unidirectional excitation and tunneling of localized surface plasmon resonance (LSPR) [23-25].

When x-polarized light is incident in the forward direction (along the −z axis), the upper grating provides the necessary wavevector to excite LSPR at the air/metal interface. The localized modes tunnel through a silver film with a thickness smaller than the LSPR penetration depth and are efficiently decoupled into the substrate by the lower grating, resulting in strong forward transmission [26].

Under backward incidence (along the z axis), the lower grating does not support wavevector conditions for LSPR excitation at the same frequency. Consequently, the incident light cannot generate LSPR and is mostly reflected by the silver film. Furthermore, near-field coupling between the upper and lower gratings suppresses backward transmission, leading to near-zero transmittivity in the reverse direction. Therefore, this kind of device can achieve a high isolation ratio exceeding 10 dB.

Electric field distribution reveals strong localized electric field in the z direction ($E_z$), particularly at longer wavelengths and with larger grating periods (e.g., $p_1$ = 900 nm), by contrast the one in the x direction ($E_x$) is negligible. Overall, the device offers a promising platform for photonic systems exhibiting asymmetric transmission.

## 3.1 Asymmetric spectra

Fig. 2 presents the forward and backward transmission spectra of the device under different upper grating periods $p_1$ = 700, 800, and 900 nm (Figs. 2**(a)-(c)**), along with the corresponding isolation ratios (Figs. 2**(d)-(f)**). These results are used to systematically analyze the AT and wavelength dependence of the structure.

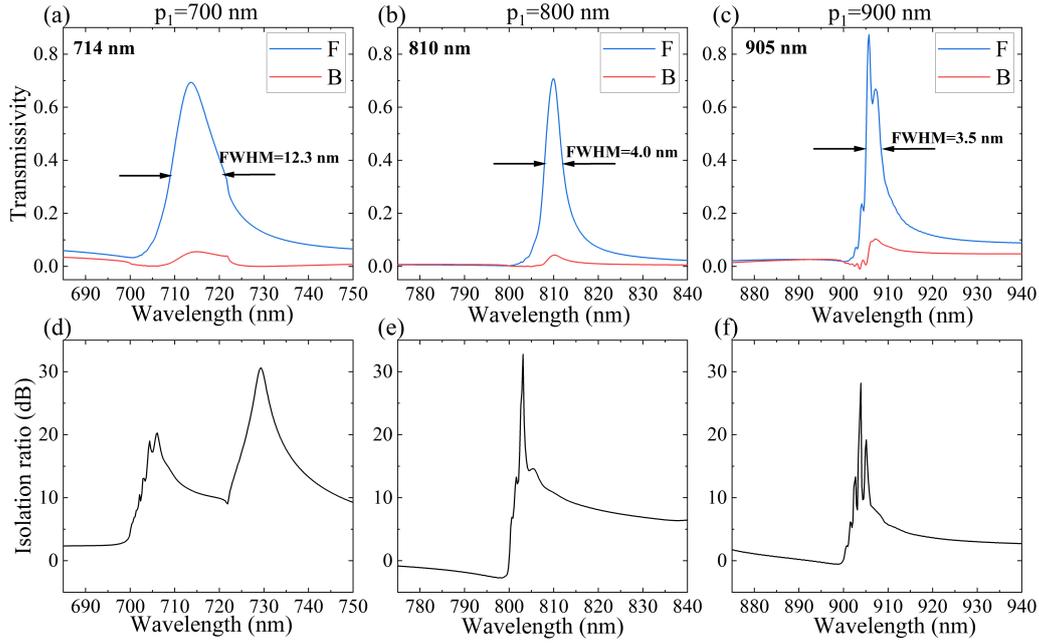

**Fig. 2.** (a)-(c) Transmission spectra of the three devices for $p_1$ = 700, 800, and 900 nm, respectively. Blue and red curves stand for forward (F) and backward (B) transmissivities, respectively. (d)-(f) Isolation ratio between forward and backward transmissivities for $p_1$ = 700, 800, and 900 nm, respectively.

Under the condition of $p_1$ = 700 nm, the transmission spectrum exhibits a clear resonance peak. For forward incidence, a main resonance peak appears near 714 nm, with a maximum transmittivity of approximately 0.69 and a full width at half maximum (FWHM) [27] of 12.3 nm, indicating a moderate spectral width. The backward transmittivity is very low, less than 0.05, suggesting strong suppression of backward-propagating light. In addition, as shown in Fig. 2**(d)**, there are two peaks in the isolation ratio at around 706 and 730 nm, respectively. These two high isolation ratios are caused by the low transmittivities of backward transmission. However, the corresponding forward transmittivities are less than 0.15, which may not be sufficient for high-efficiency photonic applications. Therefore, these two peaks are not discussed in details. The isolation ratio of $p_1$ = 700 nm reaches 11.29 dB at 714 nm, demonstrating effective optical isolation over a relatively broad wavelength range and showing potential for practical applications.

As $p_1$ increases to 800 nm, the forward resonance peak becomes significantly narrower, as shown in Fig. 2**(b)**. The peak shifts to 810 nm, with a transmittivity of around 0.71 and a FWHM of only 4.2 nm, indicating a high-quality-factor narrowband resonance. The backward transmittivity remains low, less than 0.04. The isolation ratio at 810 nm reaches 12.79 dB, achieving more efficient AT property, which is suitable for designing high-selectivity and low-crosstalk photonic components. The isolation ratio shown in Fig. 2**(e)** illustrates a higher peak at around 805 nm, which corresponds to a low transmittivity for forward transmission, and is not suitable for applications. Hence, the property at 805 nm is also not discussed in details.

When $p_1$ = 900 nm, the main resonance peak further red-shifts to 905 nm. The forward transmittivity reaches approximately 0.87, the highest among the three devices. The resonance peak becomes even narrower, with a FWHM of 3.5 nm, demonstrating excellent narrowband filtering performance. The backward transmittivity is less than 0.06, and the isolation ratio reaches 11.45 dB at 905 nm, indicating that the device still maintains strong AT property.

The transmission peak position, transmittivity, and bandwidth characteristics show clear trends across the three grating periods. The pronounced transmission asymmetry and high isolation ratio confirm the AT behavior of the device. As the grating period increases, the resonance peak redshifts, the forward transmittivity improves, and the resonance bandwidth narrows. Meanwhile, the isolation performance enhances, though the structure may become more sensitive to fabrication perturbations. These results demonstrate that tuning the grating period allows flexible control of the operating wavelength and optical response, providing an effective strategy for designing multi-band, high-contrast optical isolators.

## 3.2 Electric field modes

In order to understand the mechanisms of this AT phenomenon, the electric field is calculated and illustrated. Define M=E/E$_0$ as the ratio of the electric field intensity (E) to the one of the source (E$_0$), Figs. 3-5 show the electric field distribution in the xz plane of these three nanodevices. A brief summary is that the z-component of the electric field ($E_z$) dominates in the localized field compared with the x-component ($E_x$).

Here, we take the nanodevice with $p_1$ = 700 nm as an example to illustrate the excellent property (high isolation ratio with narrow band) and the physical mechanisms. Devices with other parameters perform similarly with the same mechanisms.

As shown in Fig. 3**(a)** and 3**(c)**, for forward transmission, due to the wave vector matching between the incident light and the LSPR mode, the incident light efficiently excites the LSPR within the upper gratings at the wavelengths of 714 nm. The LSPR mode is near the surface of both the upper ridges and the film, where $E_z$ dominates. This results in the considerable intensity of $E_x$ at the upper surface of the film. Due to the thin thickness of the film (20 nm), $E_x$ can penetrate through the film with little loss and it results in the considerable intensity of $E_x$ at the lower surface of the film. Hence, $E_x$ decouples to the dielectric ($SiO_2$) space with the help of the lower gratings. This process results in the high transmissivity of the forward transmission.

On the other hand, as shown in Fig. 3**(b)** and 3**(d)**, for backward transmission, due to the wave vector mismatching between the incident light and the LSPR mode, the electric filed of the incident light can't be efficiently localized near the lower surface of the lower ridges and the film, particularly, the intensity of $E_z$ is general near the lower surface of the lower ridges, but it is small near the lower surface of the middle film together with the small intensity of $E_x$, thus small intensity of $E_x$ at the upper surface of the middle film. When $E_x$ decouples to the free space, it results in the low transmissivity.

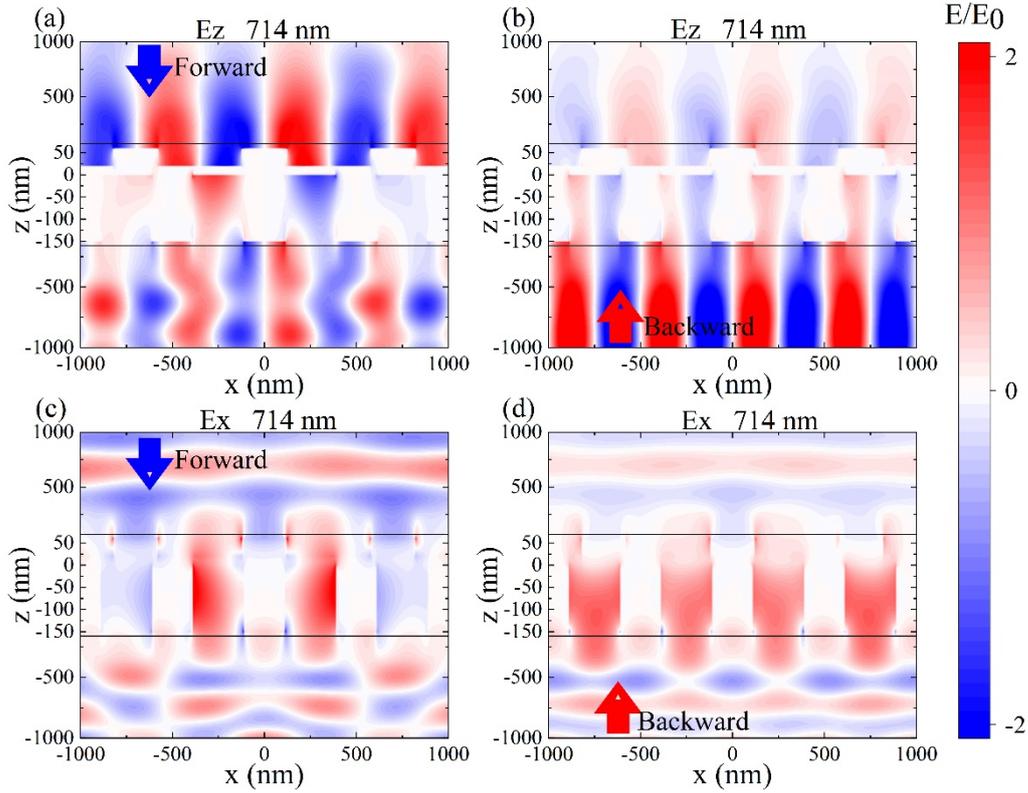

**Fig. 3.** The electric field distribution of the $p_l$ = 700 nm device in xz plane at 714 nm. (a) $E_z$ distribution under forward incidence. (b) $E_z$ distribution under backward incidence. (c) $E_x$ distribution under forward incidence. (d) $E_x$ distribution under backward incidence. The color bar represents electric field intensity relative to the one of incident light.

Due to the above analysis, the high isolation ratio of the transmissivity is obtained. The mechanism can be simply summarized using the concept of LSPR coupling and grating decoupling. That is, for forward transmission, the incident light efficiently couples with the upper gratings, excites the LSPR, resulting in the fact that $E_x$ efficiently transmits to the lower gratings, by which $E_x$ decouples to the dielectric ($SiO_2$) space; for backward transmission, the incident light can't efficiently couple with the lower gratings, resulting in the fact that $E_x$ can't efficiently transmit to the upper gratings, thus low transmissivity.

The cases of $p_l$ = 800 nm and $p_l$ = 900 nm are similar to the one of $p_l$ = 700 nm, as shown in Fig. 4 and 5, respectively. Their greatest difference lies in the intensity of the electric field of LSPR modes. The larger $p_l$ is, the larger the intensity is.

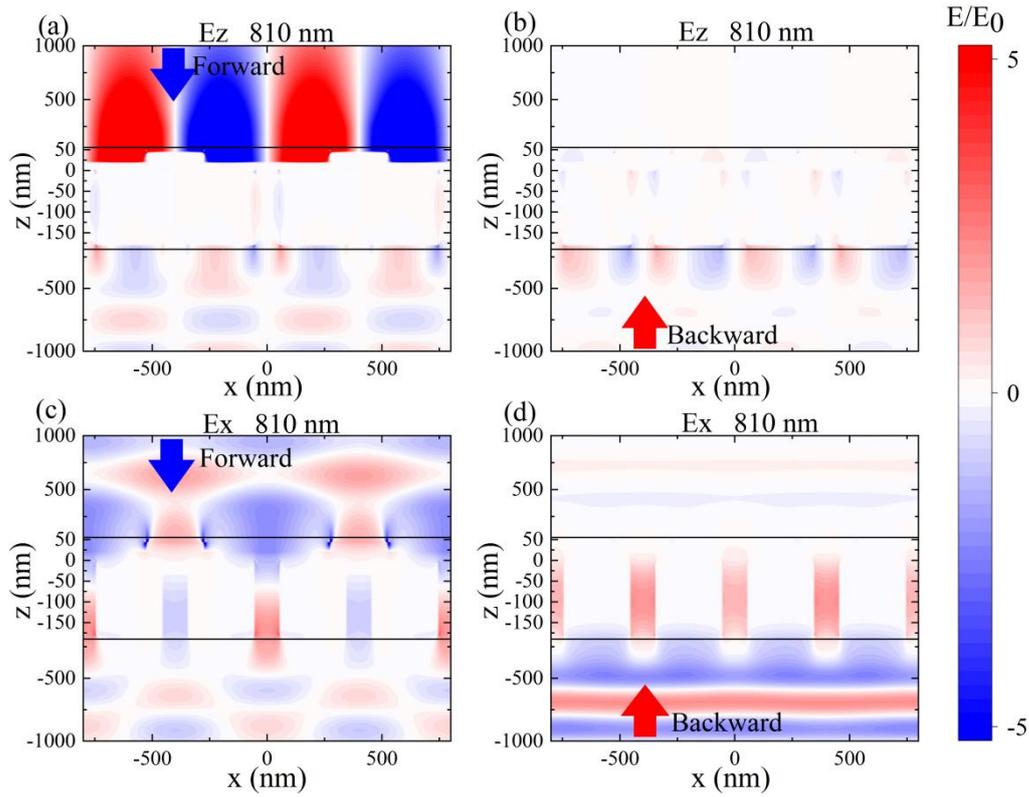

**Fig. 4.** The electric field distribution of the $p_1$ = 800 nm device in xz plane at 810 nm. (a) $E_z$ distribution under forward incidence. (b) $E_z$ distribution under backward incidence. (c) $E_x$ distribution under forward incidence. (d) $E_x$ distribution under backward incidence. The color bar represents electric field intensity relative to the one of incident light.

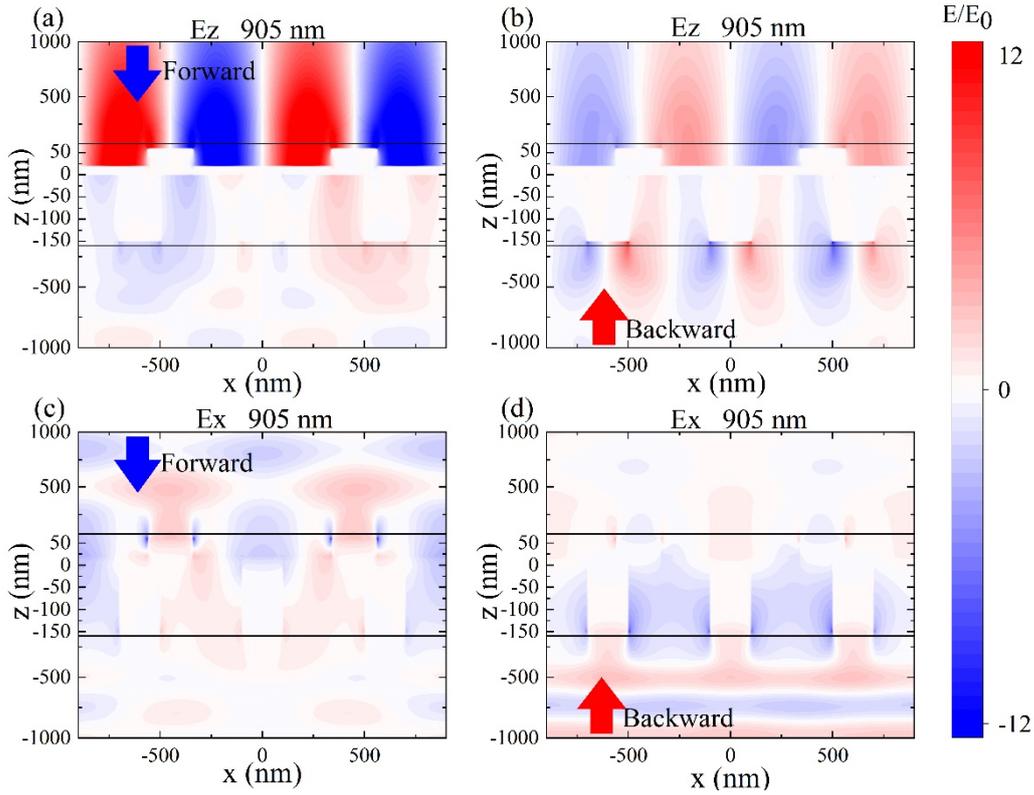

**Fig. 5.** The electric field distribution of the $p_1$ = 900 nm device in xz plane at 905 nm. (a) $E_z$ distribution under forward incidence. (b) $E_z$ distribution under backward incidence. (c) $E_x$ distribution under forward incidence. (d) $E_x$ distribution under backward incidence. The color bar represents electric field intensity relative to the one of incident light.

## *3.3 Variation of transmittivity with incidence angle*

According to the wavevector matching condition, both the resonance wavelength and the transmission efficiency are affected by variations in the incident angle ($\theta$). Fig. 6 presents the forward and backward transmission spectra under three different upper grating periods ($p_1$ = 700, 800, and 900 nm) as the incident angle increases from 0° to 10°. Figs. 6**(a)-(c)** display the forward transmission spectra for each grating period, while Figs. 6**(d)-(f)** correspond to the backward transmission spectra.

As shown in Fig. 6**(a)** and 6**(c)**, for $p_1$ = 700 nm and $p_1$ = 900 nm, the forward transmission resonance peak significantly weakens with increasing angle, and is nearly suppressed at 10°, indicating a strong angular sensitivity of the structure. In contrast, for $p_1$ = 800 nm, as shown in Fig. 6**(b)**, the overall transmission remains relatively high, and the degradation in resonance strength is much more moderate compared to the 700 nm and 800 nm cases.

Under backward illumination, as shown in Figs. **6(d)-(f)**, the transmission remains low across all angles and periods. No distinct resonance peaks are observed in the backward direction, which confirms the design principle that the lower grating cannot excite LSPR, as previously discussed.

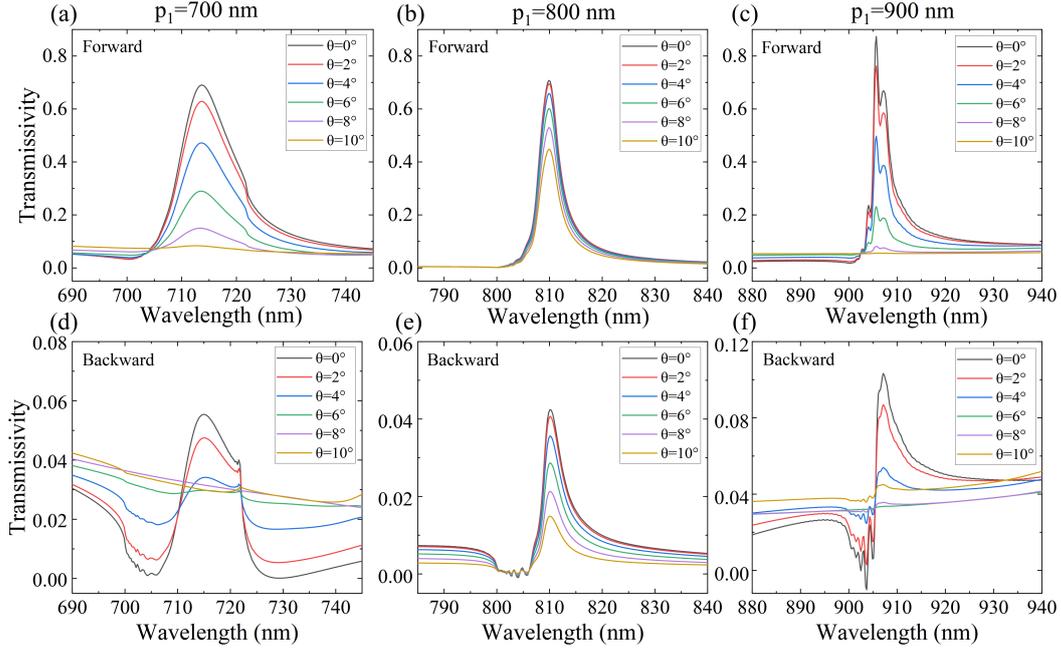

**Fig. 6.** Forward and backward transmission spectra at various incidence angles for different upper grating periods. (a)-(c) present the forward transmission spectra of the devices with $p_1$ = 700, 800, and 900 nm, respectively. (d)-(f) present the corresponding backward transmission spectra for the three devices.

### *3.4 Influence of lateral displacement of upper gratings on transmission performance*

Fig. 7 shows the forward and backward transmission spectra at $p_1$ = 800 nm, as a function of the lateral displacement (Δ) of the upper grating. It indicates that the transmission spectra are sensitive to $\Delta x$, exhibiting pronounced asymmetric transmission characteristics. Under forward illumination, the transmission peak consistently appears around a wavelength of approximately 810 nm, indicating the presence of a stable resonant mode at this wavelength. As $\Delta x$ increases from 0 nm to

200 nm, the forward transmission gradually decreases. On the other hand, the backward transmission demonstrates the same response to $\Delta x$, but maintaining low transmittivities.

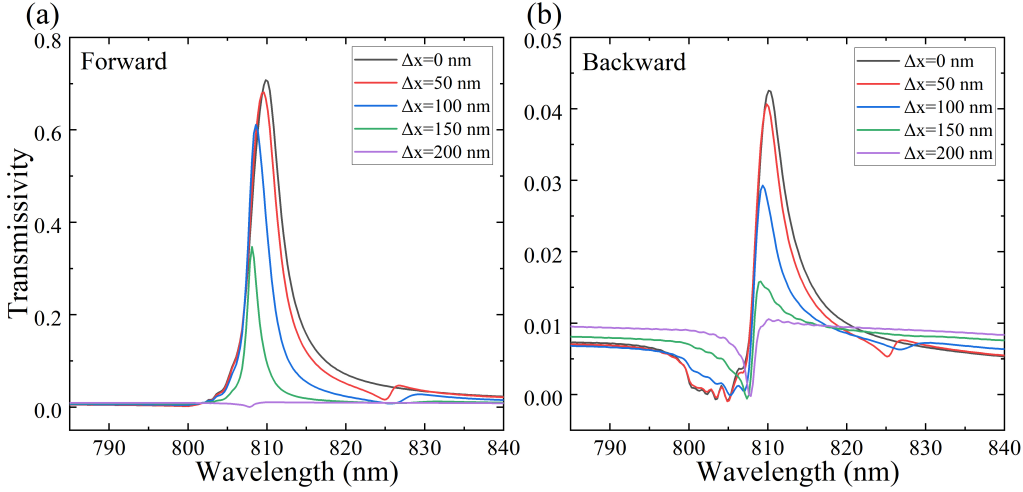

**Fig. 7.** Transmission spectrum under the condition of lateral displacement dx of the upper grating when $p_1$ = 800 nm. (a) Forward incidence; (b) Backward incidence.

It is important to note that $\Delta x$ denotes the displacement of the upper grating relative to its original position along the x-direction, which serves as a critical parameter for breaking spatial symmetry. In practical device fabrication, the precision of $\Delta x$ control is influenced by several factors, such as alignment accuracy in nanofabrication, photolithographic errors, and thermally induced displacements. These experimental uncertainties may lead to structural deviations from the theoretical design, resulting in shifts in transmission peaks or discrepancies in transmission efficiency. Therefore, particular attention must be paid to alignment and dimensional accuracy during experimental design and fabrication to ensure the realization of asymmetric transmission performance consistent with simulation results.

## 4. Conclusion

In summary, we theoretically designed three G-F-G nanostructures with different parameters, which show excellent AT properties in the red and near-infrared bands.

The devices achieve high forward transmittivity of 0.69, 0.71, and 0.87 at the wavelengths of 714 nm, 810 nm, and 905 nm, respectively, with isolation ratios all exceeding 10 dB. The principle of the high transmittivity is coupling and decoupling with the upper and lower gratings, respectively. When the light is incident from the back, the uncoupling caused by the lower grating leads in the low backward transmittivity. As a result, high isolation ratio and narrowband is achieved. Furthermore, the performance of the device is also influenced by the lateral displacement of the upper grating and variations in the incidence angle. The results indicate that the structure is sensitive to angular variations, with optimal asymmetric transmission achieved under normal incidence. As the upper grating is laterally shifted, the forward transmittivity decreases significantly, demonstrating the device's sensitivity to alignment deviations. The G-F-G structure is not only suitable for visible light band, but also has good wavelength adjustability and wide spectrum adaptability, which provides a highly feasible structural framework and physical basis for the development of multi-band integrated photonic devices, such as integrated sensors in infrared band and optical isolators.

## Acknowledgements

This work was supported by National Natural Science Foundation of China (Grant No. 12504461) and the Fundamental Research Funds for the Central Universities (Grant No. FRF-TP-20-075A1).

## Conflict of interest

Authors state no conflicts of interest.

## Data availability

The data that support the findings of this study are available upon reasonable request from the authors.